\documentclass[12pt,preprint]{aastex}
\usepackage{bm}
\usepackage{epsfig}

\begin{document}

\title{Galaxy formation and cosmic-ray acceleration in a
magnetized universe}

\author{Loukas Vlahos$^1$, Christos G. Tsagas$^2$ and Demetrios
Papadopoulos$^1$}

\affil{$^1$Department of Physics, Aristotle University of
Thessaloniki,\\ 54124 Thessaloniki, Greece}

\affil{$^2$DAMTP, Centre for Mathematical Sciences,\\ University
of Cambridge, Cambridge CB3 0WA, U.K.}


\begin{abstract}
We study the linear magneto-hydrodynamical behaviour of a
Newtonian cosmology with a viscous magnetized fluid of finite
conductivity and generalise the Jeans instability criterion. The
presence of the field favors the anisotropic collapse of the
fluid, which in turn leads to further magnetic amplification and
to an enhanced current-sheet formation in the plane normal to the
ambient magnetic field. When the currents exceed a certain
threshold, the resulting electrostatic turbulence can dramatically
amplify the resistivity of the medium (anomalous resistivity).
This could trigger strong electric fields and subsequently the
acceleration of ultra-high energy cosmic rays (UHECRs) during the
formation of protogalactic structures.
\end{abstract}


A wide variety of astrophysical and cosmological problems are
currently interpreted on the basis of gravitational instability.
The current large-scale structure and galactic evolution theories
are some of the best known examples. Relatively few of the
available studies, however, consider the role of magnetic fields,
despite the widespread presence of the latter. Magnetic fields
observed in galaxies and galaxy clusters are in energy
equipartition with the gas and the cosmic rays. The origin of
these fields, which can be astrophysical, cosmological or both,
remains an unresolved issue~\citep{K,Han}. If magnetism has a
cosmological origin, as observations of $\mu$G fields in galaxy
clusters and high redshift protogalaxies seem to suggest, it could
have affected the evolution of universe~\citep{GR,Widrow,Giovan}.
Studies of large-scale magnetic fields and their potential
implications for the formation of the observed structure have been
given by several authors (see
~\citet{T,Jacobs,Ruz,Wass,Papa,Zeldo,Adams,Barrow,
Tsagas1,Jedamzik} for a representative, though incomplete, list).
Most of the early treatments were Newtonian, with the relativistic
studies making a relatively recent appearance in the literature. A
common factor between almost all the approaches is the use of the
MHD approximation, namely the assumption that the magnetic field
is frozen into an effectively infinitely conductive cosmic medium.
With few exceptions~\citep{F,Jedamzik1}, the role of kinetic
viscosity and the possibility of finite conductivity have been
largely marginalized. Nevertheless, these aspects are essential
for putting together a comprehensive picture of the magnetic
behavior, particularly during the nonlinear regime. In this
article we consider a Newtonian expanding magnetized fluid and
assume that both viscosity and resistivity are finite. At first,
we look into the linear evolution of small inhomogeneities in the
cosmic medium and examine how the field and the fluid viscosity
affect the characteristic scales of the gravitational instability.
We then discuss the electrodynamic properties of the collapsing
fluid, the resulting magnetic amplification and the formation of
unstable current sheets. Central to our discussion is the concept
of ``anomalous resistivity'', which is triggered by electrostatic
instabilities in the plasma and can substantially reduce the
electrical conductivity of the latter. We argue that such changes
in the resistivity of the protogalactic medium will lead to the
formation of strong electric fields during the galactic collapse.
These fields can then accelerate the abundant free electrons and
ions to ultra high energies.

Let us consider an expanding, incompressible, magnetized fluid
with $p=p(\rho)$, where $p$ and $\rho$ are respectively the
pressure and the density of the matter. This medium obeys the
standard Newtonian MHD equations, which in comoving coordinates
read
\begin{eqnarray}
\frac{\partial\rho}{\partial t}&=&-3\frac{\dot{a}}{a}\rho-
\frac{1}{a}\nabla\cdot(\rho\vec{u})\,  \label{dotrho}\\
\frac{\partial\vec{u}}{\partial t}&=& -\frac{\dot{a}}{a}\vec{u}-
\frac{1}{a}(\vec{u}\cdot\nabla)\cdot\vec{u} -\frac{c_{\rm
s}^2}{a\rho}\nabla\rho+ \frac{1}{a}\nabla\phi+ \frac{1}{4\pi
a\rho}(\nabla\times\vec{B})\times\vec{B}+
\frac{\nu}{a^2\rho}\nabla^2\vec{u}  \label{dotu}\\
\nabla^2\phi&=&-4\pi Ga^2\rho\,,  \label{phi}\\
\frac{\partial\vec{B}}{\partial t}&=&-2\frac{\dot{a}}{a}\vec{B}+
\frac{1}{a}\nabla\times(\vec{u}\times\vec{B})+
\frac{\eta}{a^2}\nabla^2\vec{B}\,,  \label{dotB}\\
\nabla\cdot\vec{B}&=&0\,.  \label{gradB}
\end{eqnarray}
In the above $a$ is the cosmological scale factor, $\vec{u}$ is
the fluid peculiar velocity (with $\nabla\cdot \vec{u}=0$),
$c_{\rm s}^2={\rm d}p/{\rm d}\rho$ is the square of the sound
speed, $\phi$ is the gravitational potential, $\vec{B}$ is the
magnetic field vector, $\nu$ is the viscosity coefficient of the
medium and $\eta$ is its electric resistivity. The system
(\ref{dotrho})-(\ref{gradB}) accepts a homogeneous solution with
$\rho=\rho_0(t)\propto a^{-3}$, $\vec{B}=\vec{B}_0(t)\propto
a^{-2}$ and $\vec{u}=\vec{u}_0=0$. This solution describes a
weakly magnetized (i.e.~$B_0^2/\rho_0\ll1$) Newtonian FRW
universe, which also defines our unperturbed background.

Following Eq.~(\ref{dotu}), the magnetic effects are confined
orthogonal to $\vec{B}$ (recall that $[(\nabla\times\vec{B})
\times\vec{B}]\cdot\vec{B}=0$), which ensures that there is no
magnetic effect along the field's force lines. Given this, we
align the background magnetic field along the z-axis of an
orthonormal frame and consider its effects in the x-y plane. We do
so by perturbing Eqs.~(\ref{dotrho})-(\ref{gradB}) around the
zero-order FRW solution so that $\rho=\rho_0+\rho_1$,
$\vec{B}=\vec{B}_0+\vec{B}_1$, $\phi=\phi_0+\phi_1$ and
$\vec{u}\neq0$. Assuming wave-like perturbations
(i.e.~$\rho_1(\vec{r},t)=\tilde{\rho}_1(t)e^{i\vec{k}\cdot
\vec{r}}$, $\vec{B}_1=\tilde{\vec{B}}_1(t)e^{i\vec{k}\cdot
\vec{r}}$, etc) and using expressions
(\ref{dotrho})-(\ref{gradB}), the time derivative of
Eq.~(\ref{dotu}) gives
\begin{eqnarray}
\ddot{\vec{u}}&=&-\left(H+\frac{\nu
k^2}{a^2\rho_0}\right)\dot{\vec{u}}+ \left[8\pi
G\rho_0-\frac{k^2}{a^2}\left(c_{\rm s}^2+c_{\rm
a}^2+\frac{\nu H}{\rho_0}\right)\right]\vec{u}\nonumber\\
&{}&+i\left[\frac{c_{\rm s}^2H}{a\rho_0}\rho_1- \frac{8\pi
GaH}{k^2}\rho_1+\frac{H}{2\pi a\rho_0}
\left(\vec{B}_0\cdot\vec{B}_1\right)+\frac{k^2\eta}{4\pi
a^3\rho_0}\left(\vec{B}_0\cdot\vec{B}_1\right)\right]\vec{k}\,,
\label{ddotu}
\end{eqnarray}
where $H=\dot{a}/a$ is the Hubble parameter, $c_{\rm
a}^2=B_0^2/4\pi\rho_0$ is the Alfv\'{e}n speed squared and we have
dropped the tildas for simplicity. Also, to reduce the algebra we
have only considered perturbations orthogonal to the background
magnetic field. The real component of the above provides a wave
equation for the peculiar velocity vector, which describes a
damped oscillation. In particular,
\begin{eqnarray}
\ddot{\vec{u}}&=&-\left(H+\frac{\nu
k^2}{a^2\rho_0}\right)\dot{\vec{u}}+ \left[8\pi
G\rho_0-\frac{k^2}{a^2}\left(c_{\rm s}^2+c_{\rm a}^2+\frac{\nu
H}{\rho_0}\right)\right]\vec{u}\,,  \label{rddotu}
\end{eqnarray}
where the first term in the right shows the damping due to the
expansion and of the fluid viscosity. The latter effect is
scale-dependent and vanishes on large enough scales (i.e.~as
$k\rightarrow0$). The last term in Eq.~(\ref{rddotu}) demonstrates
the conflict between gravity on the one hand and fluid pressure
and viscosity on the other. On large scales gravity always wins
and the perturbations collapse. Small wavelength fluctuations,
however, oscillate.

Accordingly, the magnetic presence adds to the supporting effects
of pressure and viscosity only orthogonal to $\vec{B}_0$. This
means that the first scales to collapse along the magnetic field
lines are smaller than those normal to them. The two critical
wavelengths are the associated Jeans scales
\begin{eqnarray}
\lambda_\perp\simeq\sqrt{\frac{c_{\rm s}^2+c_{\rm a}^2+ \nu
H/\rho_0}{8\pi G \rho_0}} \hspace{15mm} {\rm and} \hspace{15mm}
\lambda_\parallel\simeq\sqrt{\frac{c_{\rm s}^2+\nu H/\rho_0}{8\pi
G\rho_0}}  \label{Jeans}
\end{eqnarray}
orthogonal and parallel to $\vec{B}_0$ respectively. Overall, the
magnetic presence induces a degree of anisotropy in the collapse.
Note that for a pressureless, dust-like medium the Jeans length
along $\vec{B}_0$ depends entirely on the viscosity and the Hubble
rate.

As the collapse proceeds, one expects the gradual formation of
turbulent motions within the magnetized medium. The associated
eddy viscosity is proportional to $\nu_{turb}\sim\rho_0u_1l_{mix}$
where $u_1$ is the velocity perturbation and $l_{mix}$ the
turbulent mixing length (e.g.~see~\cite{Bisk}). Assuming that
$u_1$ reaches values close to $c_{\rm a}$, that the mixing length
is a fraction of the magnetically induced Jeans length
(i.e.~$l_{mix}\ll\lambda_{\perp}$) and given the low thermal
temperature of the post-recombination universe, the characteristic
scaling on the velocities is $c_{\rm s}^2<\nu H/\rho_0<c_{\rm
a}^2$. Note that we have also adopted the typical values of
$B\leq10^{-7}$~G, $\rho_0\geq10^{-29}$~${\rm gr}\cdot{\rm
cm}^{-3}$ and $H=100\,h$~${\rm km}\cdot{\rm sec}^{-1}\cdot{\rm
Mpc}^{-1}$, where $0.4\leq h\leq1$. Then,
$\lambda_{\perp}/\lambda_{\parallel}\sim \sqrt{(c_{\rm
a}^2\rho_0)/(\nu_{turb}H)}\gg 1$, with $\lambda_{\perp}$ of the
order of a (comoving) Mpc. Following the standard structure
formation scenarios, the initial collapse of this very large
structure will be followed by successive fragmentation into
smaller scale formations with characteristic lengths
$\lambda\ll\lambda_{\perp}$. Moreover, as we will outline next,
the anisotropy of the collapse will further increase the magnetic
field trapped into the gravitating medium.

In the case of an almost spherically symmetric collapse, linear
inhomogeneities in the magnetic energy density amplify in tune
with those in the density of the matter so that $\delta
B^2\propto\delta\rho$, where $\delta B^2=B_1^2/B_0^2$ and
$\delta\rho=\rho_1/\rho_0$~\citep{TB,Tsagas}. Therefore, even
within spherical symmetry, the formation of matter condensations
in the post-recombination universe also signals the amplification
of any magnetic field that happens to be present at the time. We
have seen, however, that the generically anisotropic nature of the
field will inevitably induce some degree of anisotropy to the
collapse. Moreover, the magnetically induced anisotropy in the
collapse will backreact and affect the evolution of the field
itself. The magnetic evolution during the nonlinear regime of a
generic, non-spherical protogalactic collapse, has been considered
by a number of
authors~\citep{Z,Zeldo,Bruni,Sie,D,Dolag,Roettiger}. The
approaches are both analytical and numerical and agree that
shearing effects increase the strength of the final field, while
confining it to the protogalactic plane. Compared to the magnetic
strengths of the spherical collapse scenario, the anisotropic
increase of $B$ is stronger by at least one order of magnitude.
Thus, protogalactic structures can be endowed with magnetic fields
stronger than those previously anticipated.

So far we have seen how the magnetic presence modifies the way
gravitational collapse proceeds, by changing the overall stability
of the magnetized fluid. This in turn affects the evolution of the
field itself and can trigger a chain of nonlinear effects on
certain scales. Next we will argue that this selective
amplification of certain perturbative modes can play a important
role during the nonlinear stages of protogalactic collapse,
helping the instability to reach its saturation point. The current
induced by the total field $B$ is
\begin{equation}\label{curr}
\vec{J}=\frac{c}{4\pi}\nabla\times\vec{B}=
\frac{c}{4\pi}\alpha\vec{B}\,,
\end{equation}
where $\nabla\times\vec{B}=\alpha\vec{B}$ and $\alpha$ measures
the magnetic torsion (see~\citet{Parker} for details). Initially
$\alpha$ is small. However, the subsequent fragmentation of the
protogalactic cloud will increase $\nabla\times\vec{B}$ and
strengthen the induced current. For example, a large-scale
magnetic field with magnitude $B\sim10^{-7}$~G at the time of the
collapse will lead to $J\sim10^2\alpha$. The latter can reach
appreciable strengths for reasonable values of $\alpha$.

When the current exceeds the critical value $J_c\sim
\rho(e/m_p)c_{\rm s}$, where $c_{\rm s}\sim 10^4 \sqrt{T(^\circ
K)}$ cm/sec, $m_p$ is the ion mass and $e$ is the electron charge,
the excitation of low frequency electrostatic turbulence will
increase the resistivity of the medium by several orders of
magnitude~\citep{Galeev,Kulsrud}. Note that the typical critical
current is very small in the early post-recombination universe
(i.e.~$J_c\sim nec_s\sim10^{-10}~\textrm{statamper/cm}^2$).
Consequently, very small values of $\alpha$ will lead to $J>J_c$,
thus making the plasma electrostatically unstable. The effect,
which is known as `anomalous resistivity', can be explained
through the development of current driven electrostatic
instabilities in the plasma. The latter lead to the excitation of
waves and oscillations of different kinds. The absorption of these
waves by the ions is an additional way of transferring momentum
from the electrons to the ions, along with the usual momentum loss
from the former species to the latter. The average momentum loss
by the electron per unit time can be written as an effective
collision term in the form
$nm_e\nu_{eff}\vec{u}_{e}=-\vec{F}_{fr}$, where $F_{fr}$ is the
average friction force and $n=\rho/m_p$ the ambient number density
of the plasma particles. The friction force is proportional to the
linear growth rate of the electrostatic waves ($\gamma_k$) and the
energy of the excited waves ($W_k$). The effective collision
frequency is estimated to be $\nu_{eff}=\omega_e (W_{sat}/k_B T)$.
Then, the anomalous resistivity will be
\begin{equation}
\eta_{an}\sim\frac{\nu_{eff}}{\omega_{e}^2}
\sim\left(\frac{W_{sat}}{ k_BT}\right)\frac{1}{\omega_e}\,,
\label{eta}
\end{equation}
where $\omega_e=5.6\times 10^4 \sqrt{n}~{\rm sec}^{-1}$ is the
plasma frequency, $k_B$ the Boltzmann constant and $(W_{sat}/k_B
T)\sim 1$ is the saturated level of the electrostatic waves
~\citep{Galeev}. For certain types of current driven waves, the
anomalous resistivity is several orders of magnitude above the
classical one, as confirmed in numerous laboratory
experiments~\citep{Hamb,Yamada}.

This sudden switch to high electrical resistivity will inevitably
lead to the formation of strong electric currents and  therefore
to a fast magnetic dissipation and intense plasma heating. The
electric fields induced by the gravitational collapse will be
$E\sim c_{\rm a}B/c+\eta_{an}J_c\sim\eta_{an}J_c$, given that
$c_{\rm a}/c\ll1$. Thus, in this scenario, the gravitational
collapse of the magnetized, post-recombination cloud amplifies the
magnetic field and indirectly generates strong electric currents
localized on the protogalactic plane. The anisotropy of the
collapse enhances the local currents further and eventually drives
the resistivity towards anomalously high values. The inevitable
result is strong electric fields accelerating the abundant free
electrons. The energy gain by an electron travelling a length
$\lambda\sim\lambda_{\perp}$ is $W_{kin}\sim eE\lambda_{\perp}\sim
e\eta_{an}J_c\lambda_{\perp}$ and the relativistic factor
$\gamma=[1-(v/c)^2]^{-1}$ is given by
\begin{eqnarray}
\gamma-1&=&\frac{W_{kin}}{m_ec^2}\sim
\frac{e\eta_{an}J_c\lambda_{\perp}}{m_ec^2}\sim
\frac{e^2\eta_{an}nc_{\rm s}\lambda_{\perp}}{m_ec^2}\sim
\frac{e^2B\sqrt{T}}{m_em_pc^2\sqrt{G}\sqrt{n}}
\hspace{20mm} \nonumber\\
&\sim& 10^{11}\left(\frac{n}{10^{-4}{\rm cm}^{-3}}\right)
^{-{\textstyle{1\over2}}}
\left(\frac{T}{1^{\circ}K}\right)^{\textstyle{1\over2}}
\left(\frac{B}{10^{-7}{\rm G}}\right)\,,
\end{eqnarray}
in CGS units. Recall that $J_c\sim enc_{\rm s}$, $c_{\rm
s}\sim10^4\sqrt{T}$ and that $\lambda_\perp\sim B/nm_p\sqrt{G}$
when $c_{\rm s}^2<\nu H/\rho<c_{\rm a}^2$ (see (\ref{Jeans}b)).
Also, we have set $W_{sat}/k_B T\sim 1$ in Eq.~(\ref{eta}), which
means that $\eta_{an}\sim1/10^4\sqrt{n}$. Accordingly, the typical
energy gain by a free electron can reach extremely high values
within short timescales ($t_{acc}\sim
\lambda_\perp/c\sim10^6$~yrs), even for relatively weak magnetic
fields. Clearly, one can extend this process to proton
acceleration and show that protogalactic collapse can also source
UHECRs.

We also anticipate a few particles drifting in and out these
``primordial'' current sheets (and the associated strong
E-fields). If fragmentation has already taken place these
particles will diffuse along the different current sheets and
possibly form the observed power law distribution. The detailed
acceleration processes, however, is beyond the scope of this
article (see \citep{kaspar,Vlahos} for the diffusion of particles
in many acceleration sites).

The role of unstable currents sheets along the giant radio
galaxies on the acceleration of cosmic ray acceleration has
already been pointed out in the
literature~\citep{Kron04,Colgate,Nodes}. Particles gain and lose
(through synchrotron and inverse Compton emission) energy
continuously, by travelling at speeds close to the speed of light.
The suggestion made here is that ultra high energy cosmic-ray
acceleration and propagation may have started almost
simultaneously with the formation of galaxies through the
electrodynamic characteristics of the gravitational instability
and continue, through the same processes till today, since the
previously described instability is active on all cosmic scales.
It is also worth pointing out that the anomalous resistivity
mechanism can easily dissipate strong galactic magnetic fields, in
the form of bursty heating and particle acceleration, whenever
$\eta_{an}\nabla^2\vec{B}>\nabla\times(\vec{u}\times\vec{B})$.

The role of cosmic magnetism during the early evolution of the
first structures in our universe has been a subject of research
and debate for many decades. Most of the available studies,
however, operate within the limits of the MHD approximation, that
is they assume a highly conducting cosmic medium. As a result, the
potential large-scale implications of a magnetic presence within a
resistive environment are still relatively uncertain. In this
letter, have considered a simple scenario which starts from the
gravitational instability of a Newtonian, expanding, magnetized,
viscous and resistive fluid and discusses the implications of the
field's presence during the early phases of what one might call
the mild nonlinear protogalactic collapse. Focusing on the role of
viscosity and especially on that of electrical resistivity, we
have looked into  the electrodynamic properties of the
aforementioned gravitating medium and discussed issues such as
current sheet formation, anomalous resistivity and particle
acceleration on large scales.

We begun by outlining the ways in which a magnetic presence and a
finite fluid viscosity can alter the standard picture of
gravitational instability. We then discussed how the preferential,
anisotropic magnetic amplification will also increase the currents
on the plane perpendicular to the main axis of the collapse. These
gravitationally induced current sheets will in turn trigger
electrostatic instabilities, which can then lead to anomalous
resistivity values and subsequently to strong electric fields. We
argue that the latter can be strong enough to accelerate the free
electrons to ultra high energies.

The influence of magnetic fields on cosmic-ray propagation has
been the subject of research in the past (see~\citet{SME} and
references therein). To the best of our knowledge, however, this
is the first time that a direct connection between gravitational
instability and cosmic-ray acceleration has been suggested and
discussed. We have outlined the basic features of this connection
in a simple scenario that involves only standard Newtonian
magnetohydrodynamics. Given that we are in the post-recombination
era and that the scales of interest are well within the horizon,
we do not anticipate any general relativistic corrections. Special
relativistic effects may need to be accounted for, but not before
the particle velocities are an appreciable fraction of the light
speed. Clearly, a detailed study of the acceleration mechanism
proposed here should also consider nonlinear effects and the
possible implications of a varying electrical resistivity. The
latter has been treated as a slowly changing variable, relative to
the acceleration timescale. In any case, the key requirement for
this simple scenario to work is the presence of a magnetic field
coherent on the scale of the collapsing protogalaxy. Our
calculations argue that the required strength of this field is
comparable to those observed in high redshift protogalaxies. If
such magnetic fields are widespread, as current observations
indicate, their amplification during the nonlinear regime of
galaxy formation can trigger a range of nontrivial effects. In
this letter we suggest that these effects can include the
formation of strong large-scale current sheets and electric
fields. The latter could act as driving sources for the cosmic
rays observed in our universe today.\\

The authors would like to thank Prof.~Jan Kuijpers and Dr Heinz
Isliker for helpful discussions. GT would also like to thank the
Astronomical Observatory at the University of Thessaloniki, where
most of this work was done, for their hospitality. This project
was supported by the Greek Ministry of Education through the
PYTHAGORAS program.

\end{document}